# A Novel Multimodal biometric authentication system using Machine Learning and Blockchain

Richard Brown[1], Gueltoum Bendiab[2], Stavros Shiaeles[2], and Bogdan Ghita[1]

[1]CSCAN, University of Plymouth, PL4 8AA, Plymouth, UK

richard.brown@students.plymouth.ac.uk, bogdan.ghita@plymouth.ac.uk

[2]Cyber Security Research Group, University of Portsmouth, PO1 2UP, Portsmouth, UK

gueltoum.bendiab@port.ac.uk, sshiaeles@ieee.org

**Abstract**

Secure user authentication has become an increasingly important issue in modern society as in many consumer applications, especially financial transactions, it is extremely important to prove the identity of the user. Traditional authentication systems that rely on simple passwords, PIN numbers or tokens have many security issues, like easily guessed passwords, PIN numbers written on the back of cards, etc. Thus, biometric authentication methods that rely on physical and behavioural characteristics have been proposed as an alternative for those systems. In real-world applications, authentication systems that involve a single biometric faced many issues, especially lack of accuracy and noisy data, which boost the research community to create multibiometric systems that involve a variety of biometrics. Those systems provide better performance and higher accuracy compared to other authentication methods. However, most of them are inconvenient and requires complex interactions from the user. Thus, in this paper, we introduce a novel multimodal authentication system that relies on machine learning and blockchain, with the aim of providing a more secure, transparent, and convenient authentication mechanism. The proposed system combines four important biometrics: fingerprint, face, age, and gender. The supervised learning algorithm Decision Tree has been used to combine the results of the biometrics verification process and produce a confidence level related to the user. The initial experimental results show the efficiency and robustness of the proposed multimodal systems.

## I. Introduction

User authentication has effectively become one of the most challenges facing the digital world today. Traditional authentication methods that rely on tokens, password, and Personal Identification Number (PIN) are gradually becoming obsolete [1]. In fact, tokens and PIN/Passwords offer limited protection and can be easily lost, stolen, forgotten, guessed, or compromised [2], [3]. In this context, the last report by the World Economic Forum [4] revealed that 80% of security breaches, in 2020, are perpetrated from weak and stolen passwords. Moreover, the report affirms that, for companies, 50% of IT help desk costs are allocated to passwords resets, with average annual spend over $1 million for staffing alone [4]. These shortcomings have led to biometric authentication becoming the focus of the research community in last years. It refers to the technology that identifies and authenticate individuals in a fast and secure way through the use of unique behavioural and biological characteristics like fingerprints, hand geometry, vein, face, iris, voice, palm, DNA, etc [2]. This technology has quickly established itself as an alternative to Personal Identification Number (PIN), tokens and Passwords for various reasons [5]: Biometrics are unique for individuals and almost impossible to replicate or forge [3], which provides superior accuracy and prevent unauthorised access from those who may have the means to steal passwords or PINs [2], [5]. In addition, Biometric authentication offers convenience, accountability, and reduces the overall administrative costs by eliminating the time consuming to reset passwords [2]. In addition, they are resistant to social engineering attacks, especially phishing attacks.

Biometric technology has been considered by the research community as the most reliable and safe method for individuals authentication and several biometric systems based on common biological and behavioural characteristics (e.g. fingerprint, face, iris, handwriting, palm, Keystroke, etc.) have been developed during last decades [3], [6]. As shown in Figure.1, all biometric systems fellow the same process. First, a biometric system (e.g. fingerprint scanner, digital camera for face, etc.) is used to capture and records a specific trait of the user. The collected trait is usually analysed and translated to a template that can be stored in a database, or in a smart card that the user can carry with him [6]. This step is called enrolment. Then, each time the user authenticated for accessing the system, presented trait values are compared against those in the stored template by generating a matching score indicating the degree of similarity between the pair of biometrics data. The resulting score should be high for legitimate users and low for those from different ones. Based on the matching score also known as confidence level, genuine user is permitted access to the system and impostor is rejected. In this step, a biometric sensor is used to extract the trait being used for identification.

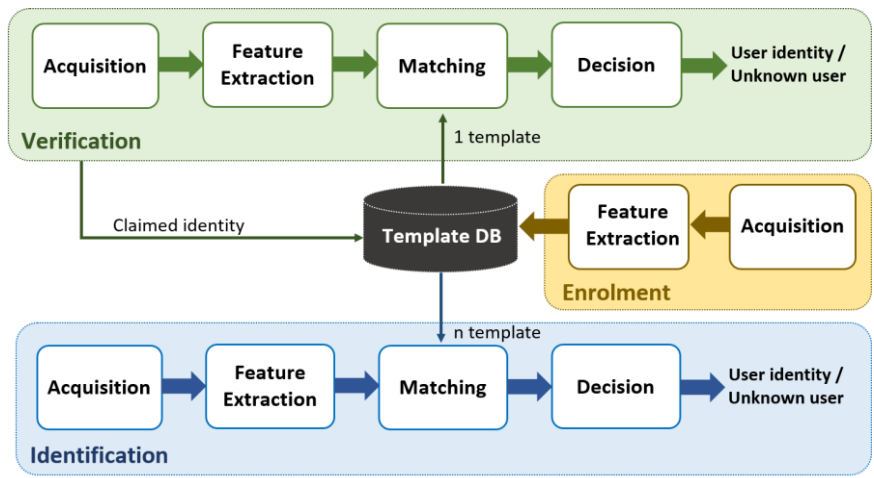

Figure.1: Biometric authentication process [7].

In real-world applications, Biometric authentication systems which involve one single biometric trait for enrolment and verification are facing a variety of problems such as lack of accuracy due to noisy data, spoof attacks, non-universality, lack of uniqueness, etc [2]. To address these limitations, there have been many attempts to create multimodal biometric systems that combine more than one physiological and/or behavioural traits for enrolment and verification [2], [6]. Usually, these systems involve a variety of biometrics that are fused, normalised and fed into a machine learning classifier to drive a decision [8]. This led to a highly accurate, secure authentication system. They also provide better performance compared with unimodal systems. However, most existing multimodal biometric systems are inconvenient and relied heavily user interaction to authenticate.

In order to fulfil the objectives of a more secure, and transparent authentication mechanism, this paper introduces a novel system for individual identity management that uses multimodal biometric authentication system with machine learning and blockchain. The multimodal biometric system combines four different biometrics for enrolment, identification, and verification. The biometrics are fingerprints, facial, and age and gender that are driven from the facial biometrics. This will increase the authentication security and overcome the limitations of unimodal systems. Based on the outputs of the biometrics verification process, the supervised learning algorithm decision tree is used to identify a confidence level related to the user. This confidence level should be high enough to allow the user accessing the protected resources on the web server (i.e. Service Provider). Whereas the blockchain is used to store user details and key data that can be used by the Service Provider (i.e. the web server) and the identity provider (i.e. BCA server) to encrypt/decrypt the user access token. Thus, the confidence in the user identity can be maintained. Further, compared to other multi modal mechanisms, the proposed system minimise the total amount of interaction required for identification and authentication.

The structure of the paper is as follows. Section 2 gives an overview of biometric authentication systems, Section 3 describes the general architecture of the proposed methodology used to ensure the user identity during time, while section 4. Finally, section 6 presents the conclusions of this work and proposes some future work and open issues.

## II. Related work

In last years, biometric authentication has become crucial, especially in security and privacy preserving applications such as, financial transactions, surveillance system, visa processing, critical environments and so on. However, due to the inherent limitations within each biometric, no single biometric method is able to achieve a high precision and reliability of individuals authentication [9]. Thus, in highly critical applications, a single biometric may not be sufficient to guarantee security, but it may be necessary to perform strong authentication by combining several biometrics [9], [10]. In this context, several multibiometric systems based on conventional physical and behavioural characteristics such as fingerprint and iris have been developed in present time. This combination of multiple biometrics is commonly referred to as multimodal biometrics authentication. In such systems, biometrics are usually combined using machine learning algorithms to drive a decision or confidence level, which will be used to either allow or deny access to the protested resources. One of the first multimodal biometrics systems was proposed by Clark NL [11], using a combination of secret knowledge and biometric-based techniques to create an Intelligent Authentication Management System (IAMS). This method uses a confidence level, which continuously updated to control the user access to protected resources. This can help in countering the increasing vulnerability of traditional knowledge-based techniques. Our system shares many aspects with this proposal in regard to the confidence level and the use of multiple authentication techniques. With the overall goal of creating a system that is robust, reliable and does not interfere with the convenience of users.

In a previous work [12], face and speaker recognition modalities are used in a serial mode, where the output of one biometric modality is used to reduce the number of possible individuals that will be checked with the second biometric. Final decision is given by the second biometric from the reduced subset of individuals. This method achieved a low False Rejection Rate (FRR) (3.9%), however, the time consumption is important compared to the fusion method. Thus, most recent multimodal biometric systems have been used the fusion method to combine the features obtained from multiple biometrics. In this context, the fusion has been applied at three different levels: at the features level, at the score matching level or at the decision level. For instance, the multibiometric approach proposed by A. Tharwat et all [10], has been explored two different fusion methods: fusion at the image level and a multilevel fusion method to combined ear and finger knuckle biometrics. The experimental results showed that the fusion at the image level can improve the overall performance of the authentication system. This method combines the ear and finger knuckle images before extracting the features that will be used by the classification module to produce an abstract value or rank. Authors highlighted that there is more than one way to successfully implement a multimodal system, although it does not cover how the user is intended to interact with the system. In addition, having a user take an image of their knuckles and ear is not a user-friendly approach.

In a recent work, J. Peng et al [13] have been proposed a multibiometric authentication system that combines four finger biometric traits: finger vein, fingerprint, finger shape and finger knuckle print. A score-level fusion method has been used to produce the overall score or confidence level of the target user based on triangular norm. The experimental results showed that the used fusion method obtained a larger distance between honest and imposter score distribution as well as achieves lower error rates. In more recent work [14], T. Joseph et al have been proposed a multimodal authentication system by fusing the feature points of fingerprint, iris and palm print biometrics. After fusing the features extracted from these biometric modalities, a secret key is generated in two stages and converted into a hash value using MD-5 hashing algorithm. A novel feature-level fusion method has been proposed by Asst. Prof. Masen M et al to combine face and iris features [15]. First, the face and iris features are extracted separately using 2D wavelet transform and 2D Gabor filters, respectively. After that, the proposed fusion method is applied by using both canonical correlation and serial concatenation. Then, the deep belief network is used for the verification process. This approach has been validated on the SDUMLAHMT database [16] and achieved an overall recognition accuracy up to 99%. However, the Equal Error Rate (EER) and fusion time are important in comparison with other systems. Many other multibiometric authentication systems have been proposed in last years by using different biometrics and different fusion methods [17], [18], [19], however, most of them are inconvenient and relied heavily user interaction to authenticate.

### III. Proposed approach

This section presents the detail about the proposed multimodal biometric system for individual's authentication using machine learning and blockchain. As shown in Figure.2, the authentication process involves three entities: the user, the Service Provider (i.e. web or resource server) and the Identity Provider (i.e. Biometric Confidence Authentication (BCA) server). The BCA server is responsible on the enrolment, identification, and verification of the user biometrics along with his level of confidence. It provides Single Sign-On (SSO) for multiple web applications. Users can monitor their confidence level and submit biometrics through a web interface (i.e. client) provide by the BCA server.

For accessing protected resources hosted by a resource server, the user must first obtain an access token from his BCA server with which he is registered. Thus, he provides his fresh biometric traits (through the sensor) together with his identity. These two pieces of information are then elaborated by the sensor and transmitted to the BCA server for verification. The BCA server queries the Database for the stored template associated with the user ID and compares it with the received one. If the templates are close enough the user will have a higher confidence level, otherwise, he will have a lower confidence level. If the obtained confidence level is lower than a predefined threshold, the user is rejected, otherwise, the BCA server generates an access token with the obtained confidence level of the user. After receiving the user access token, the resource server decrypts it by using the blockchain and check the confidence level of this user. If the confidence level is higher enough, based on its local security policy, the resource server provides the requested resource to the user, otherwise, the user request is rejected.

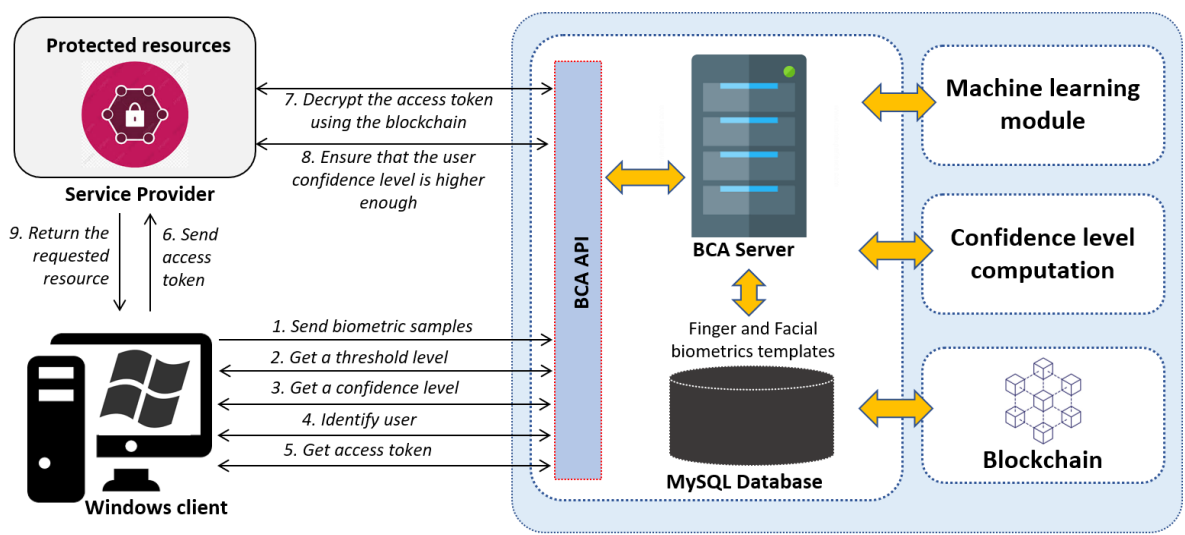

Figure.2: High level architecture of the proposed system

**1) Biometrics acquisition**

The proposed multimodal biometric system integrates four different biometrics: fingerprint, face, age, and gender. The fingerprint is the most successful and popular pattern that has been used for individuals identification and verification [6]. Fingerprints are unique and do not change in time. Their uniqueness is identified by the ridge's structures on the inner surface of a finger or a thumb. The ridges have unique local patterns, called minutiae, which have been widely used by forensic experts to match two fingerprints [20], [21]. Ridge ending and ridge bifurcation are the most commonly used patterns by the automatic fingerprint recognition systems, where a ridge ending refers to the point where a ridge ends [6],[21], while a ridge bifurcation is a point where a the ridge diverges into branch ridges [6]. Most fingerprint recognition systems use four types of fingerprint representation schemes: grayscale image, phase image, skeleton image, and minutiae [21]. In our system, The overall process of capturing the finger sample from the user takes four scans of the finger, then, the minutiae data is extracted from the Fingerprint Image Data (FID) into a template called Fingerprint Minutiae Data (FMD). The FMD is used for comparison within the system. The main reason for choosing FMD is that the original FID cannot be retrieved from the FMD as it is a one-way process.

The Facial Biometric is also known as the most distinctive key attributes for biometric authentication due to their uniqueness and robustness [22]. This technology is usually based on measurement of the facial features like mouth, eyes, nose, lips, and the face structure [22]. In this context, several techniques can be used to extract relevant features from the face image like colour analysis and neural network. In this paper, we have used the robust and fast technique Luxand FaceSDK[1] to handle the facial biometrics extraction. Luxand FaceSDK is cross-platform face detection and recognition library that provides the coordinates of over 70 facial feature points including eyes, eyebrows, mouth, nose and face contours [23]. During the enrolment phase, an image of the user is taken, and the minutiae data is extracted using Luxand SDK into a template that will be saved along with the user finger template. The facial template cannot be reversed and can only be used for comparison.

As additional biometrics, age and gender are extracted from the submitted facial image and analysed by using the Digital Persona SDK, which returns a confidence level. For instance, age result would be 'Male: 96.9999% and Female: 3.0001%'. Then, the age result is compared against the user's gender from the database and normalised in order to be consumed by the machine learning algorithm. The finger and facial templates generated in the enrolment phase are saved in a MySQL database, while the age and gender data do not need to be stored as they can be extracted on the fly. Along with those features, some other information related to the user is also saved like his identifier, name and privileges. All communications to the MySQL database are done through an ASP Web 2.0 API that was developed throughout this work.

**2) Biometrics verification & normalisation**

During the verification phase, the extracted facial and finger samples of each user are used for matching with those stored in MySQL database during the enrolment phase. The obtained similarity results are tested against a set of predefined thresholds. If the similarity values are greater than the predefined thresholds, then the comparison process returns the Boolean value "true", otherwise, it returns "false". In this step, the results from the verification and matching processes are different, some provide Boolean outputs depending on thresholds like the finger and facial biometrics matching, while other results are provided as percentages like the age and gender identification. Therefore, the obtained results should be normalised before they can be used by the machine learning module (Figure.3). For that, the obtained results for the age and the gender are also tested against predefined thresholds, if the biometrics values are greater than the predefined thresholds then, the result is set to "true", otherwise, it is set to "false".

---

[1] https://www.luxand.com/facesdk/

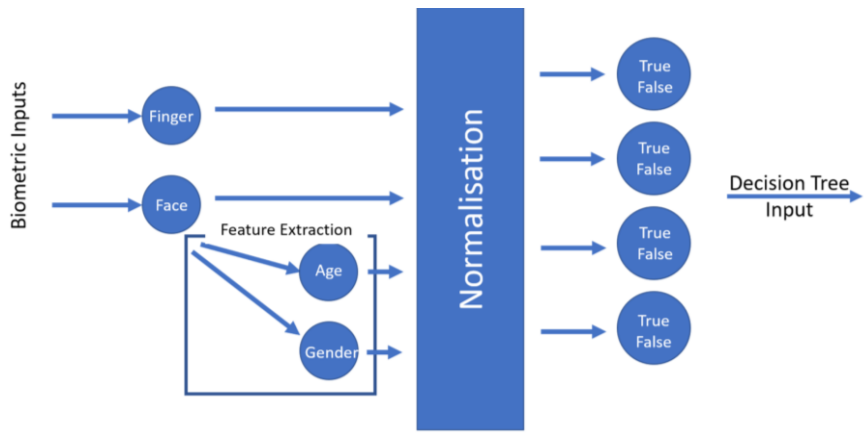

Figure.3: Normalisation Diagram

### 3) Machine learning & confidence level

The use of multimodal biometrics needs that the outcomes from multiple sources are combined to produce one result. Then the obtained result is used to figure out whether the acquired biometrics data represent a legitimate user or not. A variety of methods are available for the integration, however, in the proposed system, we will use a decision- level fusion method by using the supervised learning algorithms Decision Tree (DT) to integrate the normalised results from the four modalities and drive the confidence level related to the user. DT is a powerful and attractive approach for classification and prediction. Unlike other supervised learning algorithms, the DT has the ability to understand the given inputs and return a valuable result within a short space of time. In addition, it does not need extensive learning period compared to other methods like Neural Networks (NNs). As shown in Figure.*4*, the structure of a decision tree starts with a root node which branches out to children nodes or decision nodes, each node represents an input for the decision tree. Each children node has leaf nodes (or terminal nodes) that are the values for each of those inputs. DT predicts the value of a target variable by learning simple decision rules inferred from the data features.

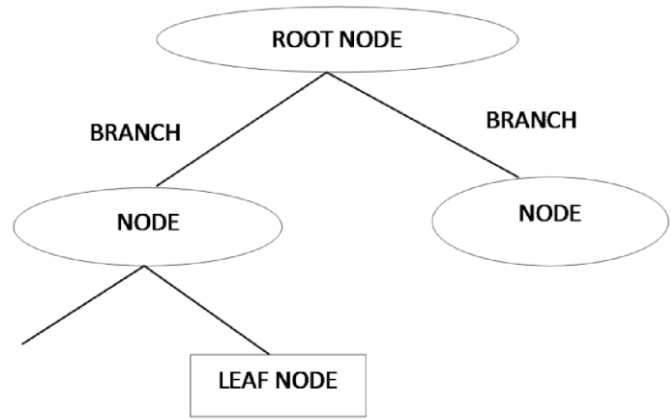

Figure.4: Basic Decision Tree Diagram [25]

As shown in Figure.*5*, first, the DT decision process is separated into two main steps. The first step is the training phase, where the DT is constructed and learned its training data (see Table-3) to understand how to interpret the inputs. In the second step, the normalised inputs will be fed into the decision tree to drive a decision or a confidence percentage. This value represents the confidence level associated with the user, which will be used to update the user's confidence by adjusting the obtained confidence value directly via a connection to MySQL database. The value of this attribute is then used to produce the confidence level of the target user and decide whether to give him access to the protected resources or not. If the user does not obtain the required confidence level, he cannot access the protected resources hosted by the webserver. The required confidence level is defined by the resource server based on its local security policy. For the decision process, the biometrics were weighted as follow, finger and facial samples are weighted at 40 % and age and gender are weighted at 10 %.

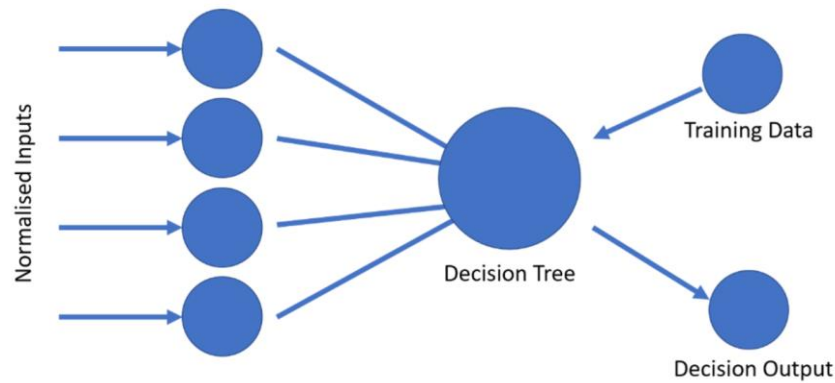

Figure.5: Decision Tree Architecture

**4) Blockchain**

The emergence of Bitcoin[2] has inspired fresh thinking about the benefits of applying the blockchain technology to the areas of identity management due to its distributed, fault-tolerant and transparent structure that can guarantee trust among untrusted parties without relying on specific trusted, central authority [26]. The blockchain is an encrypted ledger that is distributed and replicated among the members of a peer-to-peer network, containing a linear sequence of chained blocks that are capable of generating trust without external trusted authority, thus renders it difficult to compromise the integrity of their records without being noticed by the entire network, and making massive data breaches very difficult, if not theoretically impossible [26], [27]. All these characteristics were contributed to the rise of many promising and innovative blockchain-based identity management solutions [28].

In this work, the blockchain consists of a number of participating resource servers and BCA servers and it is used as a public shared ledger to store user details and key data in the form of transactions. The key data stored in transactions is used by the BCA servers and untrusted resource servers to encrypt/decrypt the users 'access tokens and prove their authenticity. Each transaction contains the user's ID, timestamp, Key, start date, end date, and previous block hash. A new transaction is created and added to the chain when a new user is enrolled in the system. The BCA server does the mining for the block on creation to avoid clients having to do intensive processing to preserve the user experience. The resource server can use the blockchain to decrypt the access token sent by the user and check if his confidence level is higher than the predefined threshold.

## IV. Experimental results

In this section, we present the experiments carried out over the proposed identity management system in order to demonstrate its effectiveness and reliability.

### 1. Experimental setup

As shown in Figure. 6, the simulation experiments were performed on a client/server environment based on the Microsoft. NET technology, where each entity is deployed in a separate VM. The overall process of capturing the finger samples from the users is performed using the Fingerprint Reader Software "DigitalPersona 4500", while the facial samples have been captured using "Luxand SDK" library. Then, the templates generated from the enrolment phase are stored in MySQL database, where a BLOB field is created for each type of templates. All communications to the database were done through an ASP Web 2.0 API that was developed in this work. The machine learning component was implemented using the Accord Framework for .NET[3]. This framework allows a smooth implementation of the DT learning algorithm on the BCA Server compared to PyTorch. Unlike other learning algorithms, DT does not require intensive training and make quick decisions, which is very important for the performance of the authentication system. The blockchain has been implemented using Microsoft. NET framework.

---

[2] https://bitcoin.org/fr/

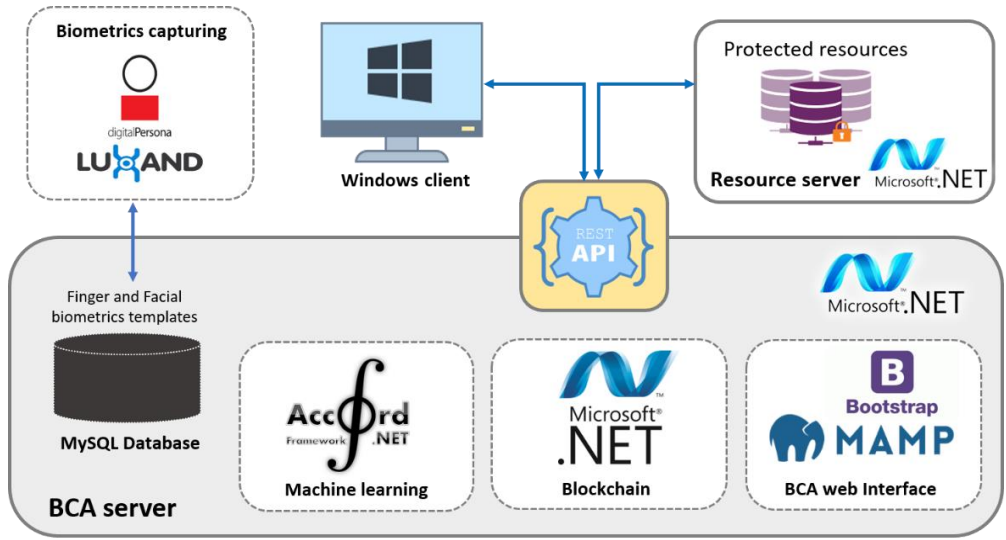

*Figure. 6: Implemented technologies for the simulation experiments*

In this work, a user-friendly GUI has been added to provide a dashboard that can be used by administrators for controlling users, viewing analytical data, and managing predefined thresholds on the BCA system (see Figure.7). The GUI has been implemented using the Bootstrap framework[4] which provides a quick and customised design of more professional web interface with HTML5.

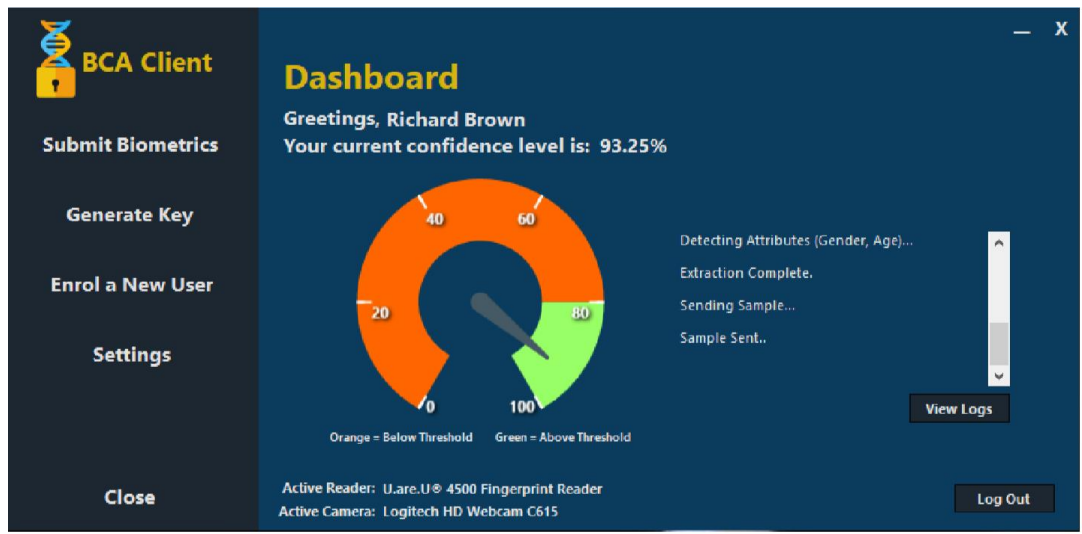

*Figure.7: BCA server dashboard*

---

[3] http://accord-framework.net/
[4] https://getbootstrap.com/

## 2. Discussion and Results

### 2.1. Thresholds values and accuracy

For testing the efficiency and performance of the proposed authentication system, several experiments have been carried out to identify the suitable thresholds values that can offer a balance between the system performance and accuracy. Therefore, the thresholds values for the fingerprint and facial biometrics are set using the recommended SDK threshold values, which provides numerical values of the threshold for setting the False Positive Identification Rates (FPIRs) and False Acceptance Rate (FAR). FAR (also referred to as "Type II error") is the percentage of frauds that were incorrectly recognised over the total tested samples [24]. FAC is a serios biometric security error as it gives unauthorised users access to the system. Thus, it must be reduced to the minimum possible.

FPIR (also referred to as "Type I error") is the ratio of the test cases that are classified above a threshold "T" (True) over the total tested samples. The threshold "T" is used to classify a test case to be either a correct (true or positive) case or false (negative) case. If the case is below a threshold "T" then it is classified false (negative) and if it is above threshold "T" it is classified true (positive). Table Error*! No text of specified style in document.*-1 shows the thresholds for the fingerprint biometric and their relationship with the FPIRs.

Table **Error! No text of specified style in document.**-1: Finger Thresholds and their Relation to FP Identification Rates

| Thresholds "T" | Corresponding FPIRs | Expected number of FP identifications | Numeric value of the threshold "T" |
|---|---|---|---|
| .001*maxint | .1% | 1 in 1,000 | 2147483 |
| .0001*maxint | .01% | 1 in 10,000 | 214748 |
| .00001*maxint | .001% | 1 in 100,000 | 21474 |
| 1.0e-6*maxint | .0001% | 1 in 1,000.000 | 2147 |

In our system, we select the threshold value 21474, which gives FPIR of .001% along with an expected number of FP of 1 in 100,000 identifications. The rate of FPI can be reduced more, however, this can increase the authentication problems for legitimate users. For example, if their fingers are sweaty, greasy, or slightly damaged, this will prevent them from accessing the system due to the authentication failures.

For the facial thresholds, the False Acceptance Rate (FAR), when the system incorrectly identifying an unauthorized person, depends on the threshold value and the total memory limit set on the capture. The higher the memory limit, the higher the false acceptance rate. FAR is also considered the most serious of biometric security errors as it may give impostors access to the system. Table-2 shows the relationship between the thresholds, the memory limit, and the FAR. In our system, the memory limit is set to 1024 MO for a facial template with a threshold of 0.992 and a FAR around 0.0002%, which is considered acceptable as it is used in conjunction with other biometrics. With the proposed thresholds the authentication system achieved high accuracy values ranging from 0.99% to 100%, with FAR of 0.0002% for facial biometric and FPIR of 0.001% for fingerprint.

Table-2: Facial Thresholds and their relationship with FARs and Memory Limits

| Thresholds "T" | Memory limits (MO) | | | | | |
|---|---|---|---|---|---|---|
| | 350 | 700 | 1750 | 3500 | 5250 | 7500 |
| 0.992,000 | 0.000,041 | 0.000,114 | 0.000,703 | 0.001,287 | 0.001,938 | 0.002,475 |
| 0.993,141 | 0.000,035 | 0.000,104 | 0.000,519 | 0.001,099 | 0.001,744 | 0.002,010 |
| 0.994,283 | 0.000,031 | 0.000,095 | 0.000,462 | 0.000,882 | 0.001,377 | 0.001,574 |
| 0.995,424 | 0.000,013 | 0.000,054 | 0.000,304 | 0.000,646 | 0.000,953 | 0.001,212 |
| 0.996,566 | 0.000,013 | 0.000,045 | 0.000,211 | 0.000,458 | 0.000,700 | 0.000,812 |

| 0.997,707 | 0.000,009 | 0.000,038 | 0.000,146 | 0.000,314 | 0.000,435 | 0.000,566 |
| --- | --- | --- | --- | --- | --- | --- |
| 0.988,849 | 0.000,006 | 0.000,006 | 0.000,066 | 0.000,161 | 0.000,276 | 0.000,327 |
| 0.999,990 | 0.000,000 | 0.000,003 | 0.000,000 | 0.000,003 | 0.000,013 | 0.000,022 |

## 2.2. Confidence level

Several experiments were performed on the proposed authentication system by considering a real-world case in which the right permissions and identity of six users have been checked. The biometrics matching results are processed by the trained learning algorithm in order to provide the overall confidence level of the user at each authentication transaction. The overall threshold for the confidence level is set to 80%. This value means that at one of the less weighted biometrics (Ager or gender) is not true. Table-3 presents the data used for training the DT learning algorithm.

Table-3: Training data for the DT learning algorithm

| Finger | Face | Gender | Age | Confidence Level |
| --- | --- | --- | --- | --- |
| true | True | true | True | 100.00% |
| True | True | true | False | 80.00% |
| true | True | false | True | 80.00% |
| True | True | false | False | 70.00% |
| True | False | true | True | 60.00% |
| True | False | true | False | 50.00% |
| True | False | false | True | 50.00% |
| True | False | false | False | 40.00% |
| False | True | true | True | 60.00% |
| False | True | true | False | 50.00% |
| False | True | false | True | 50.00% |
| False | False | true | True | 20.00% |
| False | False | false | True | 10.00% |
| False | False | false | False | 0.00% |

The graph in Figure.8 shows the evolution of confidence level values over time for one user. During this period of time, the user sent its biometric samples to the authentication system in more than 100 transactions, with different biometric samples of this user. From the obtained results, it is noticed that the confidence level values of the user are changing as expected, where the confidence level of this user stay over the threshold (from 82% to 86%) for all good samples and has been dropped below the threshold (78.6%) with bad biometrics samples. The same observations have been achieved for all users.

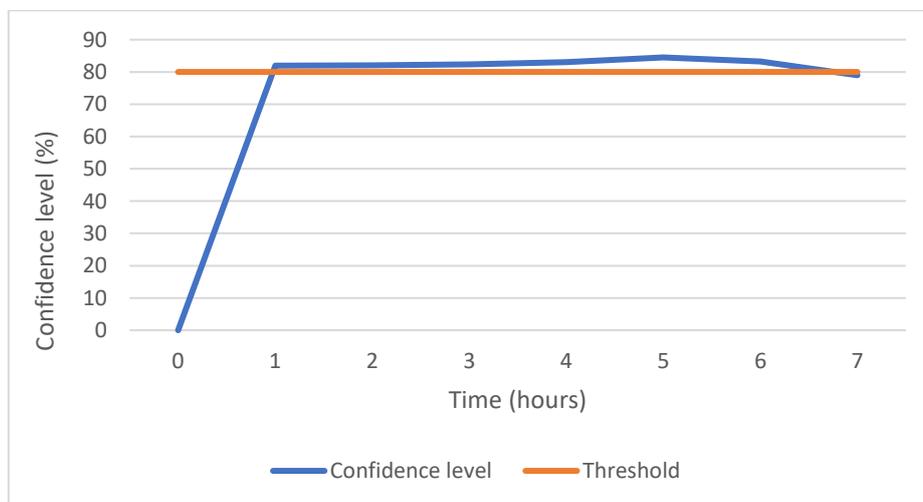

Figure.8: Confidence level values over time for one user

## V. Conclusions and perspectives

In this paper, we proposed a multimodal authentication system using fingerprint, face, age, and gender biometrics. The proposed scheme employs the DT learning algorithm to compute the user's confidence levels based on the submitted biometrics. The later is then used by the proposed system to authenticate the users. It should be higher than a predefined threshold in order the user can have access to the system. The effectiveness of the proposed system has been justified using a real-world case in which the right permissions and identity of six users have been checked, with a set of more than 100 biometric samples for each user. The samples are classified from bad to good samples.

The experiments results showed the system behaved as expected, where the good samples obtained higher confidence values and the bad samples obtained lower confidence levels. However, more experiments are needed to confirm the efficiency of the proposed approach, thus, we intend to extend this work with more experiments on large data sates from real-world as well as testing the robustness of this system against different security attacks. We also intend to further reduce the time for biometric submission by fully automating this process and minimise the user interactions, which makes the proposed system more suitable for real-time applications where computation speed is crucial.